\documentclass[aps,prl,twocolumn,superscriptaddress,showpacs,letter]{revtex4}
\usepackage[dvips]{graphicx}
\usepackage{times}

\begin{document}
\title{Plasmon excitations in planar sodium clusters}

\author{Bao-Ji Wang} 
\affiliation{Key Laboratory of Advanced Microstructured Materials, MOE,
Department of Physics, Tongji University, 1239 Siping Road, Shanghai 200092, P.
R. China}
\affiliation{College of Physics and Chemistry, Henan Polytechnic University, 2001 Shiji Road, Jiaozuo 454000, P. R. China}

\author{San-Huang Ke} 
\email[Corresponding author, E-mail: ]{shke@tongji.edu.cn}
\affiliation{Key Laboratory of Advanced Microstructured Materials, MOE, Department of Physics, 
Tongji University, 1239 Siping Road, Shanghai 200092, P. R. China}
\affiliation{Beijing Computational Science Research Center, 3 Heqing Road,
Beijing 100084, P. R. China}

\begin{abstract}
The collective electronic excitation in planar sodium clusters is studied by
time-dependent density functional theory calculations. The formation and
development of the resonances in photoabsorption spectra are investigated in
terms of the shape and size of the 2-dimensional (2-D) systems. The nature of
these resonances is revealed by the frequency-resolved induced charge densities
present on a real-space grid. For long double chains, the excitation is similar
to that in long single atomic chains, showing longitudinal modes, end and
central transverse modes. However, for 2-D planes consisting of ($n\!\times\!n$) atoms with $n$ up to 16, new 2-D characteristic modes emerge regardless of the symmetries considered. For a kick parallel to the plane, besides the equivalent end mode, mixed modes with contrary polarity occur, while for an impulse perpendicular to the plane there will be corner, side center, bulk center, and circuit modes.  Our calculation reveals the importance of dimensionality for plasmon excitation and how it evolves from 1-D to 2-D.

\end{abstract}

\maketitle

\section{Introduction}

Plasmons are collective electronic excitations in a metal caused by an electromagnetic field, which is generally related to the surface plasmon resonance (SPR) \cite{Mustafa2011-197601, Jain2007-107}. The SPR effect in 10-1000 nm wavelength region can be explained as a collective oscillation of free electrons responding to the electromagnetic perturbation. However, for systems consisting of only tens or hundreds of atoms and of a size bridging the atomic and condensed phase scales, they feature several extraordinary properties which are neither exactly the same as the Mie plasmon in  a classical metal sphere \cite{Mie1908-377} nor like the compressional bulk plasmon.  The difference is related to the existence of discrete energy levels in small nanoparticles or atomic clusters, in contrast to the situation of continuous energy bands present in solids \cite{Lian2009-174701}.

The ability of modern technology, like lithography techniques,  to produce nanoparticle arrays with uniform size and shape as well as regular interparticle separations has made it possible to investigate quantitatively the far- and near-field interactions of the surface plasmons in these periodic metal-nanoparticle structures \cite{Waele2007-2004,Hicks2005-1065,Halas2011-3913,Koenderink2007-201403, Lamprecht2000-4721,Henzie2007-549, Wei2004-1067,Sung2008-4091, Haynes2003-7337, Salerno2005-543, Maier2003-205402} 
On the other hand, some artificial atomic structures \cite{Crommie1993-218, Nilius2002-1853, Nazin2003-216110, Nazin2003-77} with different sizes and shapes were also successfully fabricated by making use of the technique of scanning tunneling microscope (STM), which have been used as ideal prototpye structures for  theoretical studies of collective electron oscillations (plasmon), because the size and shape of these artificial structures are tunable down to the precision of a single atom. In particular, the successful assembly of linear Au atomic chains on NiAl (110) surface with the STM technique \cite{Nilius2002-1853} has arisen an extensive interest in studying the plasmon excitations in small 1-dimensional (1-D) atomic chains \cite{Zangwill1980-1561,Lian2009-174701, Yan2007-216602, Yan2008-235413,Nayyar2011}.
  
Theoretically, it is important to investigate the collective optical response of these short 1-D systems directly from the first principles because Mie's theory \cite{Mie1908-377} is not applicable.  In this regards the time-dependent density functional theory (TDDFT) \cite{Zangwill1980-1561} has been approved as a powerful tool due to its ability of dealing with excited states and its high computational efficiency \cite{Lehtovaara2011-154104}. By using TDDFT calculation in the frequency domain Lian {\it et. al.} \cite{Lian2009-174701} studied the possibility of plasmon excitations in linear Au atomic chains with different lengths and found  that the collective plasmon modes in the absorption spectrum arises when the number of atoms is larger than $\sim$ 10. Yuan {\it et al.} \cite{Yan2007-216602,Yan2008-235413} studied the electronic excitations and the nature of end and central plasmon resonances in linear atomic chains of simple and noble 
metals by using a real-space and real-time TDDFT method. They found that in case of Ag atomic chains the $d$ electrons can lead to decreased intensity and energy of the transverse modes but do not affect much the longitudinal modes. Nayyar {\it et al.} studied the optical properties of short Au atomic chains doped with different transition metal atoms (Ni, Rh and Fe) by using the TDDFT approach and found that a weak doping with some transition metal atoms may lead to the generation of local plasmonic modes and, therefore, the transition metal atoms can be used to tune the optical properties of nanostructures \cite{Nayyar2011}. 

As is well known, dimensionality is one of the most important material-defining parameters: The same chemical components can exhibit dramatically different properties depending on whether the atoms are arranged in a 0-, 1-, 2- or 3-D manner. Despite the several recent theoretical studies on the plasmon excitations in the 1-D atomic chains, few attention, to our best knowledge, has been paid to the plasmonic properties of 2-D atomic planes which are now possible to fabricate experimentally and may show
nontrivial different behavior from the corresponding 1-D chains. In this work, we investigate the electronic excitations of 2-D  sodium atomic planes with different sizes and shapes by performing TDDFT
calculation in the time domain. The photoabsorption spectra due to the collective excitations caused by an impulse parallel or perpendicular to the atomic plane are studied. The nature of each resonance peak in the spectra is revealed by the frequency-resolved induced charge density present on a real-space grid. Compared to the related single Na atomic chains which have been studied previously as prototype metallic chains \cite{Yan2007-216602,Yan2008-235413}, the plasmon excitations in these 2-D systems exhibit unique characteristic features in the spatial distribution of their induced charge density, showing some completely different 2-D behavior. Our result reveals the importance of dimensionality in the formation and development of plasmon excitations in low-dimensional metallic structures and how it will evolve from 1-D to 2-D. 

\begin{figure}[b]
\includegraphics[width=7.0cm,clip]{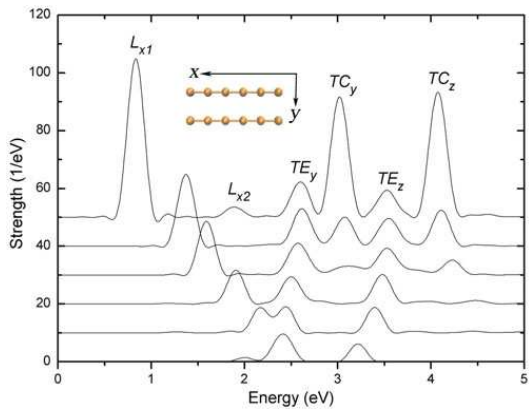}
\caption{\label{fig:na2ac-as} The dipole response of sodium double atomic chains ($2\!\times\!n$)
with a varying length $n$. The interatomic distances in the $x$ and $y$ directions are 
$a$=2.89\AA\ and $b$=4.08\AA, respectively. The series of spectra correspond, counting
from the bottom, to $n\!=\!$ 2, 3, 4, 6, 8, and 18, respectively.}
\end{figure}
 
\section{Computation}

The Na atomic planes studied in the present work are assemblies of the corresponding single Na atomic chains with a fixed intra-chain atomic separation and inter-chain spacing which are chosen to be $a\!=\!2.89\,$\AA\  and $b\!=\!4.08\,$\AA, respectively, being the same as in the experiment of building linear atomic chains on a NiAl surface \cite{Nilius2002-1853} where the spacings are determined by the lattice constants of the surface. We label an atomic plane by $(m\!\times\!n)$ where $m$ is the number of single atomic chains and $n$ is the number of atoms in each chain. The ratio of $n/m$ defines the shape of the plane. In this work, we consider two cases: double chains $(2\!\times\!n)$ with $n$= 2, 3, 4, 6, 8, and 18, and square planes $(n\!\times\!n)$ with $n$= 2, 3, 4, 5, 6, 7, 8, 12, 16. All the 2-D structures are placed in the $xy$-plane with the single atomic chains being along the $x$-axis (see the insets of Figs.\ref{fig:na2ac-as} and \ref{fig:naplaneab-as}). 

Our calculations are carried out with a real-space and real-time TDDFT program \textit{Octopus}\cite{Marques2003-60}, in which the Na ion is described by norm-conserving pseudopotentials and the local density approxiamtion (LDA) is used for the electron exchange and correlation for both the ground-state and excited-state calculations. A grid in real space which is defined by assigning a sphere around each atom with a radius of $6\,$\AA\ and a uniform mesh grid of $0.3\,$\AA\  is adopted to describe the wavefunction and charge density. To obtain the excitation spectrum, the system is impulsed from its initial ground state with a very short delta-function-like perturbation, and then the time-dependent Kohn-Sham equation is evolved in real space and real time for a certain period of time. Specifically, an electronic wave packet is evolved for 10,000 time steps with each being 0.003 $\hbar$/\,eV long. After the real-time propagation, the photoabsorption spectrum is  extracted by Fourier transforming the time-dependent dipole strength. Furthermore, a 3-D image of the frequency-resolved induced density distribution in real space is then obtained for each resonance in the spectrum by Fourier transforming the time series of the total induced charge density at each resonance frequency for every real-space mesh grid and then is visualized by a graphic software \cite{Humphrey1996-33}.

\begin{figure}[b] 
\includegraphics[width=6.5cm,clip]{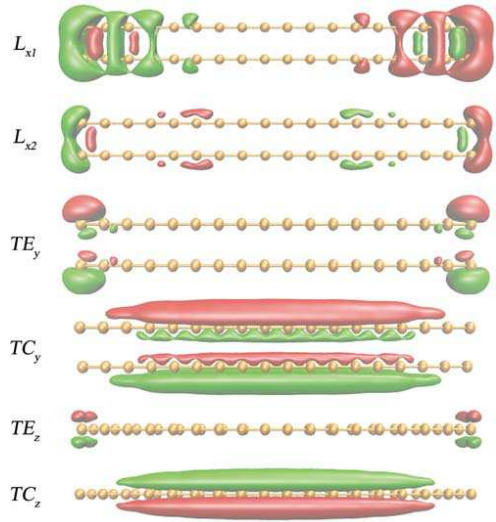}
\caption{\label{fig:na2ac-ft}
Fourier transforms of the induced densities at the frequencies of the plasmon resonances indicated on the left for the double chain ($2\!\times\!18$), as shown in Fig.~\ref{fig:na2ac-as}. 
Green and red isosurfaces denote the negative and positive densities, respectively.
}
\end{figure}

\section{Result and Discussion}

\subsection{Double chains}

Let us start with the double chains.  Fig.~\ref{fig:na2ac-as} shows the evolution of the (optical absorption) dipole strength as a function of the length (i.e., the number of atoms in each single chain ($n$)). One can see that as the double chain becomes long enough ($n \gtrsim 6$), its two transverse modes gradually split into four branches with two low-energy resonances impulsed along $y$-axis and two high-energy resonances caused by a kick along $z$-axis (perpendicular to the double-chain plane). That is to say, compared with the results of single Na chains given by a previous calculation (Fig. 1 of Ref.~\cite{Yan2007-216602}), one transverse mode of a single chain now split into two due to the asymmetry between the $y$ and $z$ directions, but with similar evolution in energy as the length of the chain increases. 
The low-lying longitudinal $L_{x1}$ mode behaviors similarly as in the case of single chain: It redshifts in energy as the double chain becomes longer, which results from the accumulation of collectivity in the dipole oscillation. Its intensity increases near linearly with the chain length, which can arise from the reduction of the energy gaps involved in the dipole excitation \cite{Yan2008-235413}. 

The nature of these resonances can be understood by looking at the induced charge densities 
at the resonance frequencies. Fig.~\ref{fig:na2ac-ft} shows the frequency-resolved induced densities for the double chain with $n$=18. It can be seen that  $L_{x1}$ is a single longitudinal mode and $L_{x2}$ is a multipolar longitudinal mode which may be resulted from the longitudinal quantization of plasmon, as previously found in long single chains \cite{Yan2008-235413} .
On the other hand, the $TE_z$ ($TE_y$) and $TC_z$ ($TC_y$) peaks 
in Fig.~\ref{fig:na2ac-as} correspond to the end and central transverse modes, respectively, when the kick is along $z$-axis ($y$-axis). The frequencies of $TE_y$ and $TC_y$ modes are lower than those of the $TE_z$ and $TC_z$ modes, respectively,  because one more atom is added in the $y$ direction.
In both cases, the $TE$ mode has a lower excitation energy than the $TC$ mode.
Such a spatial separation of the transverse excitations resembles the surface and bulk plasmons of solids and thin films, where the surface (end) plasmons have lower energies than the bulk (central) plasmon. 
Overall, our calculation shows that the collective excitation in long double chains (with a large length-width ratio) is similar to that in long single chains although the transverse modes are now split. 

\begin{figure}[t]
\includegraphics[width=7.0cm,clip]{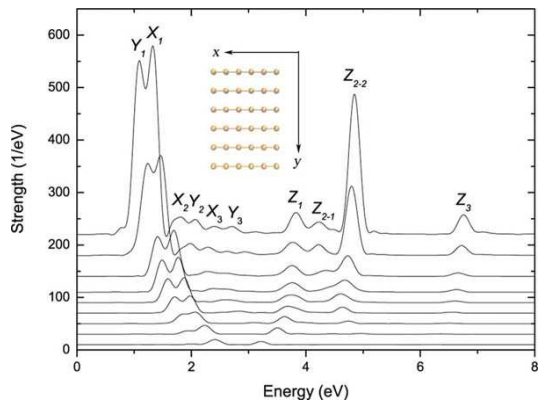}
\caption{\label{fig:naplaneab-as}
The dipole response of the asymmetrical square sodium planes ($n\!\times\!n$) with a varying size $n$. The interatomic distances in the $x$ and $y$ directions are $a$=2.89\AA\ and $b$=4.08\AA, respectively. The series of spectra correspond, counting from the bottom, to $n\!=\!$ 2, 3, 4,5, 6,7, 8 ,12, and 16.}
\end{figure} 

\begin{figure}[t]
\includegraphics[width=7.0cm,clip]{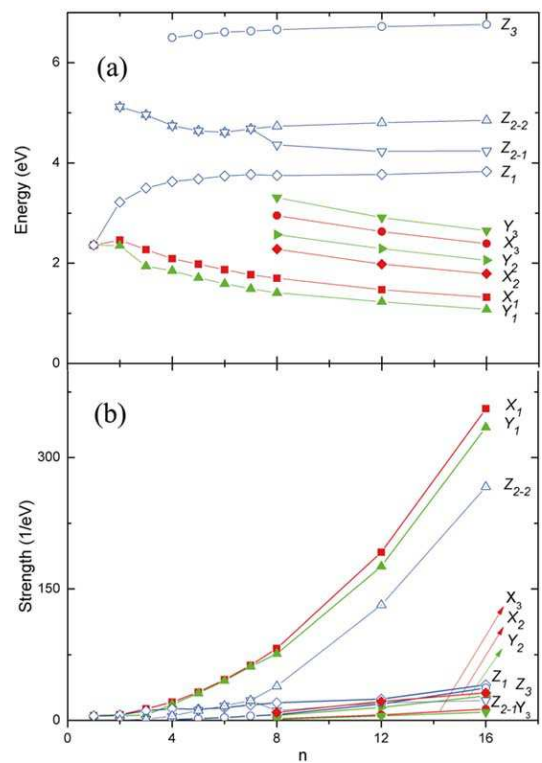}
\caption{\label{fig:naplaneab-es} The excitation energy (a) and dipole strength
(b) as a function of the size $n$ of the asymmetrical square sodium planes ($n\!\times\!n$), for the resonance peaks shown in Fig.~\ref{fig:naplaneab-as}.}
\end{figure}

\subsection{Asymmetrical square planes}

Next we consider the case of square planes, $(n\!\times\!n)$, with different sizes $n\!=\!2, 3, 4, 5, 6, 7, 8, 12, 16$, and plot their excitation spectra in Fig.~\ref{fig:naplaneab-as}. 
It is obvious that starting from the same smallest cluster $(2\!\times\!2)$ the evolution of the resonance peaks in energy for the square planes is totally different from that for the single or double chains. When the impulse is along the $x$- or $y$-axis the resulting low-energy resonances $X_1$ and $Y_1$ are split because of the asymmetry ($a \neq b$). One should note that the increase in the interatomic separation ($b > a$) will lower the corresponding resonance frequency ($Y_1$ is lower in energy than $X_1$). This can be understood by considering the fact that the nonlocal $s$ electrons can travel a bit longer distance with a larger interatomic separation in the $y$ direction. When the square plane becomes large enough ($n \gtrsim 8$) some other higher-energy resonances ($X_2$, $Y_2$, $X_3$, and $Y_3$ in Fig.~\ref{fig:naplaneab-as}) begin to emerge. These resonances behave differently from $X_1$ and $Y_1$: Now the $Y$ peaks are higher in energy than the $X$ peaks, indicating their oscillation modes are different (see discussion later).
A common trend for all these plasmon resonances in the larger planes is that their energy redshifts and their strength increases near linearly with the increased size ($n$) of the plane (see Fig.~\ref{fig:naplaneab-es}). The linear redshift can be understood by considering the reduction of the energy gap as the cluster becomes larger. The increase in strength can be ascribed to the accumulation of collectivity in the excitation as more electrons participate in the collective oscillation.  

When the dipole excitation is along the $z$-axis, there is only one mode ($Z_1$)
visible for the two smallest clusters (i.e., ($2\!\times\!2$) and
($3\!\times\!3$), see Fig.~\ref{fig:naplaneab-as}). As the atomic plane becomes
larger, more high-energy modes gradually emerge. The first one is $Z_2$ and it
dominates in intensity starting from ($6\!\times\!6$). When the plane is larger
than ($8\!\times\!8$) this resonance splits into two peaks ($Z_{2-1}$ and
$Z_{2-2}$) and then the $Z_{2-2}$ increases its intensity rapidly and develops
into a dominant plasmon resonance (a 2-D central mode, as discussed later).
Starting from ($5\!\times\!5$) another high-energy mode ($Z_3$) emerges. As shown in Fig.~\ref{fig:naplaneab-es}, the intensities of all these $Z$ plasmon modes in the larger planes increase near linearly as the size of the plane gets larger because of the accumulation of collectivity. However, the change in their excitation energies is much smaller compared to the case of the $X$ and $Y$ modes because the dipole oscillation is perpendicular to the atomic plane and therefore is affected much less by the size of the plane, as similarly found in the case of single atomic chains \cite{Nie2009-295203}. 

\begin{figure}[t]
\includegraphics[width=7.5cm,clip]{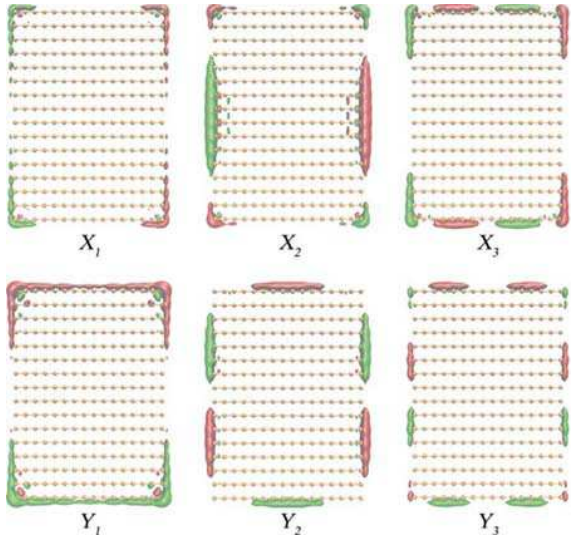}   
\caption{\label{fig:naplaneab-ft1} 
Fourier transforms of the induced densities at the frequencies of the plasmon resonances excited along the $x$ and $y$ directions for the asymmetrical square plane ($16\!\times\!16$), as shown in Fig.~\ref{fig:naplaneab-as}.
}
\end{figure}

\begin{figure}[tb]                                
\includegraphics[width=7.5cm,clip]{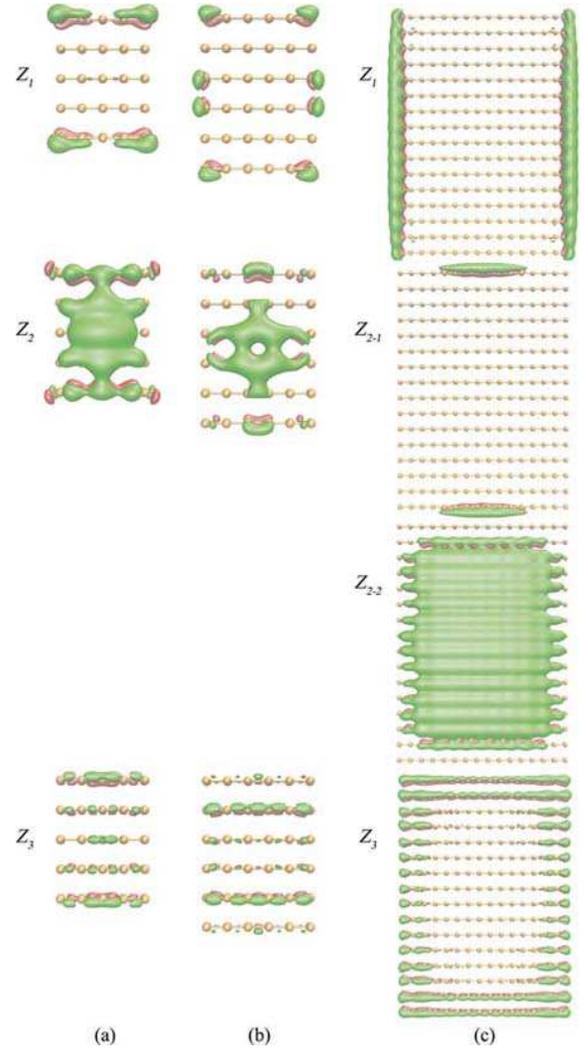}
\caption{\label{fig:naplaneab-ft2} 
Fourier transforms of the induced densities at the frequencies of the plasmon resonances excited along the $z$ direction for (a) ($5\!\times\!5$), (b) ($6\!\times\!6$), and (c) ($16\!\times\!16$) asymmetrical square planes, as shown in Fig.~\ref{fig:naplaneab-as}. 
} 
\end{figure}

To understand the nature of the $X$ and $Y$ plasmon resonances in the square Na planes we plot their frequency-resolved induced charge densities in Fig.~\ref{fig:naplaneab-ft1} for the ($16\!\times\!16$) system. 
For the $X_1$ and $Y_1$ resonances, the induced density is mainly distributed around the four corners of the plane, which is somewhat similar to the $TE$ mode in single chains. This is understandable because the 2-D plane can be viewed as a zonal chain along the direction perpendicular to the impulse direction. However, the asymmetry between the $x$- and $y$-direction leads to slightly different excitation energies between the $X_1$ and $Y_1$ modes as shown in Fig.~\ref{fig:naplaneab-as}. The $Y_1$ has a slightly lower energy because the nonlocal $s$ electrons can travel a bit longer distance in the $y$ direction.
Differing from the $X_1$ and $Y_1$ modes, the $X_2$, $Y_2$, $X_3$, and $Y_3$ modes have a mixed density distribution. For $X_2$, besides the equivalent center (EC) mode, an equivalent end (EE) mode with the opposite polarity also exists simultaneously. For $Y_2$, there is a EC mode and a multipolar end (ME) mode with the opposite polarity existing simultaneously. Similarly, $X_3$ is EE+ME and $Y_3$ is ME+ME. Because the two simultaneous modes have opposite polarity the total dipole strength of these mixed modes is much smaller than $X_1$ and $Y_1$. The different plasmon excitation also lead to a different order of their excitation energies, for instance, $Y_2$ has a higher energy than $X_2$.
%
Note that the asymmetry between the $x$- and $y$-direction is the reason for the nonequivalence between the $X$ and $Y$ plasmon resonances, as can be seen clearly in  Fig.~\ref{fig:naplaneab-ft1}: upper panel vs lower panel.


When an impulse perpendicular to the Na plane is imposed, the frequency-resolved
induced charge densities of the resulting $Z$ plasmon modes are plotted in
Fig.~\ref{fig:naplaneab-ft2} for three sizes of the plane (($5\!\times\!5$),
($6\!\times\!6$), and ($16\!\times\!16$)) to show the evolution of these modes.
One can see that when the Na plane is small (($5\!\times\!5$) and
($6\!\times\!6$))  the density distribution is quite irregular but evolves to form regular patterns when the Na plane becomes large enough ($16\!\times\!16$): $Z_1$ turns out to be an equivalent end mode with its density located around the longer sides of the plane; $Z_2$ evolves gradually to form two modes with their density distributed around the center of the shorter sides ($Z_{2-1}$) and in the bulk center ($Z_{2-2}$); $Z_3$ evolves eventually into a circuit mode. Like in the case of single and double atomic chains, here the end/surface modes ($Z_1$ and $Z_{2-1}$) have lower excitation energies than the central/bulk mode ($Z_{2-2}$). Note that the induced density distributions from all the $Z$ plasmon modes are highly complementary in space. 

\begin{figure}[tb]
\includegraphics[width=7.0cm,clip]{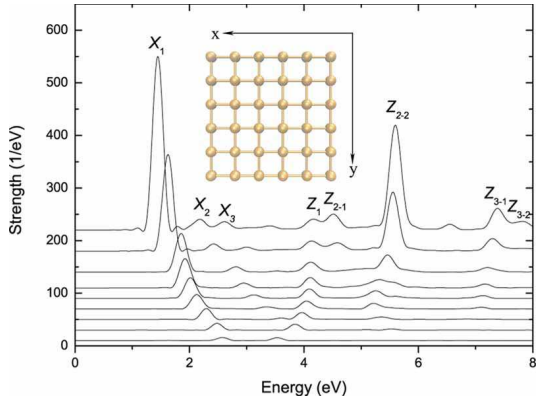}
\caption{\label{fig:naplaneaa-as}
The dipole response of the symmetrical square planes ($n\!\times\!n$) with a varying size $n$. The interatomic distances in the $x$ and $y$ directions are $a\!=\!b$=2.89\AA, respectively. The series of spectra correspond, counting from the bottom, to $n\!=\!$ 2, 3, 4,5, 6,7, 8 ,12, and 16.
}
\end{figure}

\begin{figure}[tb]
\includegraphics[width=7.5cm,clip]{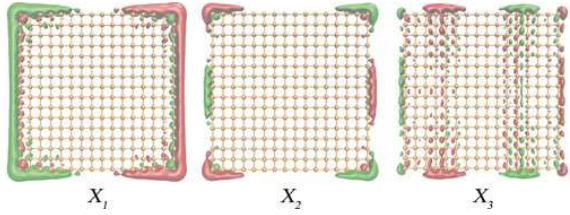}
\caption{\label{fig:naplaneaa-ft1}
Fourier transforms of the induced densities at the frequencies of the plasmon resonances excited along the $x$ and $y$ directions for the symmetrical ($16\!\times\!16$) square plane, as shown in Fig.~\ref{fig:naplaneaa-as}.
}
\end{figure}

\begin{figure}[tb]
\includegraphics[width=7.5cm,clip]{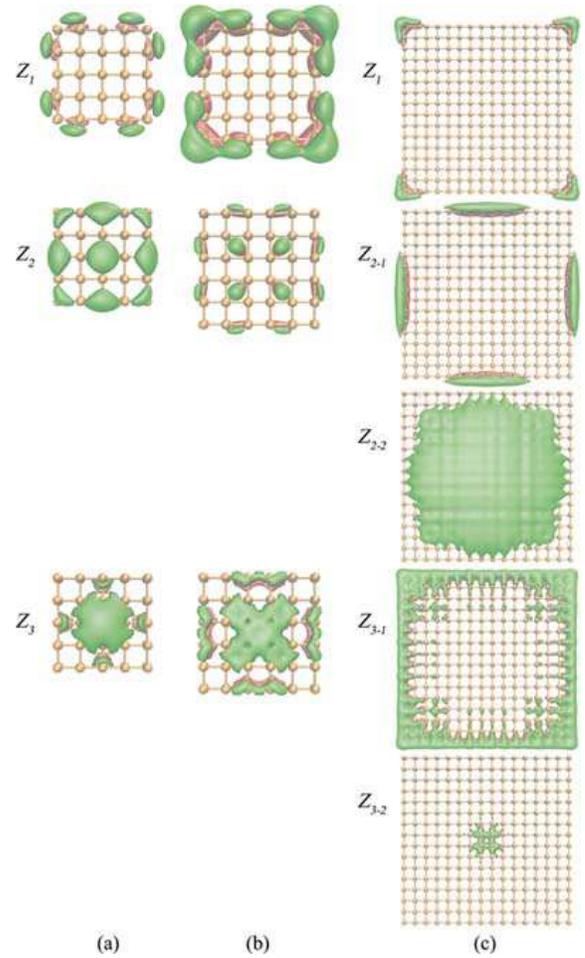}
\caption{\label{fig:naplaneaa-ft}
Fourier transforms of the induced densities at the frequencies of the plasmon resonances excited along the $z$ direction for (a) ($5\!\times\!5$), (b) ($6\!\times\!6$), and (c) ($16\!\times\!16$) symmetrical square planes, as shown in Fig.~\ref{fig:naplaneaa-as}. 
}
\end{figure}

\subsection{Symmetrical square planes}

From the density distribution of the $Z_1$ or $Z_{2-1}$ plasmon mode one can see clearly the significant effect from the asymmetry of the plane ($a \neq b$). To have a clearer picture of this effect and to see more generally the plasmonic behavior in 2-D systems, we also perform calculations for the symmetrical Na planes with $a\!=\!b\!=\!2.89\,$\AA. The photoabsorption spectra are given in Fig.~\ref{fig:naplaneaa-as} for the systems ($n\!\times\!n$), $n\!=\!2, 3, 4, 5, 6, 7, 8, 12, 16$. 
By comparing these results with those of the asymmetrical planes (Fig.~\ref{fig:naplaneab-as} ) one can see that now the $X$ and corresponding $Y$ peaks merge into single ones (i.e., $X_1$, $X_2$, and $X_3$ peaks in Fig.~\ref{fig:naplaneaa-as}) because of the symmetry. The energy of the $X_1$ (symmetrical) is also quite close to the $X_1$ (asymmetrical) because now lattice constants in both $x$ and $y$ directions are the same as the lattice constant in the $x$ direction of the asymmetrical systems. 
By looking at the frequency-resolved induced densities shown in
Fig.~\ref{fig:naplaneaa-ft1} one can also see that the density distributions of
the $X_{1,2,3}$ modes in the symmetrical ($16\!\times\!16$) plane are very close to those of the $X_{1,2,3}$ in the asymmetrical  one but are remarkably different from those of the $Y_{1,2,3}$.



On the other hand, this change in symmetry does not, however, affect qualitatively the $Z$ modes: the overall shape of the spectra almost remains the same (Fig.~\ref{fig:naplaneaa-as} vs Fig.~\ref{fig:naplaneab-as}) except for the remarkable blueshifts in Fig.~\ref{fig:naplaneaa-as} and a new high-energy resonance ($Z_{3-2}$) emerging in ($16\!\times\!16$).
To see the nature of these remarkably blueshifted $Z$ modes, we plot their
frequency-resolved induced density in Fig.~\ref{fig:naplaneaa-ft} for the
symmetrical ($5\!\times\!5$),  ($6\!\times\!6$), ($16\!\times\!16$) planes. One
can see that now the charge distribution is symmetrical in the $x$ and $y$
directions for all the $Z$ resonances because of the symmetry which also leads
to a quite different evolution and the eventual pattern in the large
($16\!\times\!16$) plane for each $Z$ mode.  Now $Z_1$ evolves into a pure corner mode instead of the longer-side mode. $Z_2$ develops finally into two modes, a side-center mode and a strong bulk center mode which is similar to that in the asymmetrical case and is the most dominant mode in dipole strength in both cases. $Z_3$ evolves eventually into a circuit mode and a weak central mode. Like in the case of asymmetry, here the induced densities from the different modes are highly complementary in space.  

Our calculations show that the plasmon excitation in 2-D atomic structures is sensitive to the structural symmetry. In the present case, a change of symmetry in the $xy$-plane will affect qualitatively the $X$ and $Y$ resonances but will not have a qualitative effect on the $Z$ resonances although the quantitative effects can be significant, like the remarkable blueshifts and the changes in the detailed pattern of collective charge oscillation.

\section{Summary}

We have studied the collective electronic excitations in planar sodium clusters by performing the time-dependent density functional theory calculations in the time domain. Two kinds of systems are considered: double chains and square atomic planes. The formation and development of the collective resonances in their photoabsorption spectra are investigated as a function of their size. The nature of these plasmon modes is revealed by the frequency-resolved induced charge densities present on a real-space grid. 

For the double chains with a large length-width ratio, the longitudinal modes are similar to those in single chains, while the transverse modes show two pairs of bimodal structure because of the lower symmetry, which can be still assigned as the end and central modes, respectively, just as in the case of single atomic chains. However, for larger square atomic planes consisting of ($n\!\times\!n$) atoms with $n$ up to 16, new 2-D characteristic modes emerge. For a kick parallel to the plane, besides the equivalent end mode, the induced density also shows mixed modes with contrary polarity. On the other hand, for an impulse perpendicular to the plane the induced density shows 2-D characteristic corner, side center, bulk center, and circuit modes.  

The structural symmetry of an atomic plane plays a significant role in its plasmon excitation. Changing a atomic plane from being in-plane symmetrical to asymmetrical will lead to splitting of the degenerated in-plane modes. It will also cause remarkable changes in the nature and excitation energies of the perpendicular modes, but will not change qualitatively the overall shape of the perpendicular photoabsorption spectra. 

Our calculations reveal the importance of dimensionality in the formation and development of plasmon excitation in low-dimensional metallic structures and how it will evolve from 1-D to 2-D.

\section{Acknowledgments}

This work was supported by the MOST 973 Project under Grant No. 2011CB922204 and by the Shanghai Pujiang Program under Grant No. 10PJ1410000 as well as by the National Natural Science Foundation of China under Grant No. 11174220.


\end{document}